\documentclass[preprint2]{aastex}
\usepackage{subfigmat}
\usepackage{subfigure}
\usepackage{lineno} 
\usepackage{amsmath} 

\slugcomment{Draft version 5.3 (18th January 2020)}

\shorttitle{GRB170405A with multi-wavelength observations}
\shortauthors{Arimoto et al.}

\begin{document}

\title{Physical origin of GeV emission in the early phase
of GRB 170405A: Clues from emission onsets with multi-wavelength
observations \\}


\author{Makoto Arimoto\altaffilmark{1}, Katsuaki Asano\altaffilmark{2}, Yutaro Tachibana\altaffilmark{3}  and Magnus Axelsson\altaffilmark{4}\altaffilmark{5} }
\email{arimoto@se.kanazawa-u.ac.jp}



\altaffiltext{1}{Faculty of Mathematics and Physics, Institute of Science and
Engineering, Kanazawa University, Kakuma, Kanazawa, Ishikawa 920-1192, Japan}
\altaffiltext{2}{Institute for Cosmic Ray Research, The University of Tokyo, 5-1-5 Kashiwanoha, Kashiwa, Chiba,  277-8582, Japan}
\altaffiltext{3}{Tokyo Institute of Technology, 2-12-1 Ookayama, Meguro-ku, Tokyo, 152-8551, Japan}
\altaffiltext{4}{Department of Physics and Oskar Klein Center for Cosmoparticle Physics, Stockholm University, 106 91 Stockholm, Sweden}
\altaffiltext{5}{Department of Physics, KTH Royal Institute of Technology, AlbaNova, SE-106 91 Stockholm, Sweden}

\begin{abstract}
The origin of GeV emission from the early epoch of gamma-ray bursts (GRBs) is a widely discussed issue.
The long gamma-ray burst GRB 170405A, observed by the {\it Fermi} Gamma-ray Space Telescope showed high-energy emission delayed by $\sim$20\,s with respect to the X-ray emission, followed by temporally fading gamma-ray emission lasting for $\sim$1,000 s, as commonly observed in high-energy GRBs. In addition, a high-energy spectral cutoff at $\sim$50 MeV was detected in the prompt emission phase.
If this feature is caused by pair-production opacity, the bulk Lorentz factor of the GRB ejecta can be estimated to be $\Gamma_{\rm bulk}$ =  170--420.
Simultaneously with {\it Fermi}, GRB 170405A was observed by {\it Swift}/Burst Alert Telescope (BAT) and X-ray telescope (XRT), and a clear optical onset was detected $\sim$200 s after the burst by {\it Swift}/UltraViolet Optical Telescope (UVOT). By coupling the deceleration time to the derived bulk Lorentz factor, the deceleration time was found to correspond to the delayed onset in the optical band.
While the delayed onset in the optical band is evidence that this emission had an external shock origin, the temporally extended emission in the GeV band before the optical onset is hard to reconcile with the standard synchrotron emission from the same external shock. This may imply that the common feature of GeV emission with a power-law decay does not necessarily have the same origin of the optical afterglow in all {\it Fermi}/LAT GRBs, particularly in their early epoch.  Another emission mechanism to explain the GeV emission in GRB 170405A can be required such as  an internal-shock or inverse Compton emission. 

\end{abstract}


\keywords{gamma-rays: bursts --- gamma-rays: observations --- gamma-ray bursts: individual(GRB170405A)}



\section{Introduction}

High-energy gamma-ray emission from gamma-ray bursts (GRBs) is commonly observed by the {\it Fermi} Gamma-ray Space Telescope with its two onboard instruments: the Gamma-ray Burst Monitor (GBM), which has a coverage of 8 keV--40 MeV  \citep{2009ApJ...702..791M}; and the Large Area Telescope (LAT), which has a coverage of 20 MeV--300 GeV \citep{2009ApJ...697.1071A}. Several observations have revealed that almost all GRBs show some common features such as (1) delayed GeV emission with respect to the X-ray--MeV emission in the GBM band and (2) temporally extended emission lasting $\geq$1,000 s and starting from the middle of the prompt-emission phase \citep{2013ApJS..209...11A, 2FLGC}.
This implies that the temporal behavior observed in the GeV band has different origins from that in the X-ray band, where the standard synchrotron shock model \cite[see][for reviews]{2004RvMP...76.1143P} has mostly succeeded in explaining the typical behaviors of the prompt emission and the afterglow with the internal-shock and external-shock scenarios, respectively.
The model is also successful in the optical band, where most observations were in the afterglow phase and the temporal and spectral behaviors seen can be described by the external-shock origin.
In the GeV band, however, the simple standard model cannot be easily applied to the spectral and temporal behaviors, and several alternative scenarios have been proposed: leptonic models of inverse Compton emission in an internal shock \citep[e.g.,][]{2010ApJ...720.1008C, 2011ApJ...739..103A},   a reverse shock \citep[e.g.,][]{2016ApJ...831...22F,2017ApJ...848...94F}, or a $e^{+/-}$ pair-loaded blast wave that originates from the MeV radiation front of GRBs in prompt-emission phase \citep[e.g.,][]{2014ApJ...788...36B,2015ApJ...813...63H}, synchrotron emission in an external shock \citep[e.g.,][]{2010MNRAS.409..226K, 2010MNRAS.403..926G}, and hadronic models including proton--synchrotron \citep{2010ApJ...724L.109R} or proton-cascade processes \citep{2010ApJ...725L.121A}.

Among these models,  synchrotron emission from an external shock might represent a better choice to explain the observed features of both the delayed emission in the prompt phase and the subsequent, temporally extended emission in the afterglow phase, although the model has several challenges. 
    In the context of the external shock model, the delayed onset of the initial bright gamma-ray emission can be explained by the deceleration timescale of the external shock \citep{2010MNRAS.409..226K, 2010MNRAS.403..926G, 2010A&A...510L...7G}.
 In this case, a very early onset time in the GeV band during the prompt emission for some LAT GRBs ($t_{\rm onset} \sim$ a few seconds) requires a large bulk Lorentz factor \citep[$\Gamma$ $\sim$ 600--2000:][]{2010MNRAS.403..926G, 2013ApJS..209...11A,2018A&A...609A.112G}, which is close to the theoretical limit expected from thermal acceleration in the standard fireball model \citep{2004RvMP...76.1143P,2010ApJ...709..525L}. 
  Although the temporal delay can be described by the external-shock model, it is difficult for the GeV flux in the initial bright pulse to be fully explained by this model, and other contributions might be needed \citep[e.g.,][]{2014ApJ...788...36B,2017ApJ...848...94F}.  
 Nevertheless, one of the simple approaches to test the external shock model for explaining the delayed onset is to independently measure the bulk Lorentz factor.
  For example, the spectral feature of a high-energy cutoff caused by the optical thickness of electron--positron pair creation \citep{2011ApJ...729..114A, 2015ApJ...806..194T, 2018ApJ...864..163V} is a suitably direct estimator of the bulk Lorentz factor. 
 
For the extended emission, the synchrotron emission model from the external forward shock can be easily applied to explain the observed power-law decay emissions \citep{2010MNRAS.409..226K, 2010MNRAS.403..926G, 2010A&A...510L...7G,2014MNRAS.443.3578N,2016ApJ...831...22F, 2017ApJ...848...94F}.
 Even though the GeV light curve follows a simple power-law decay, all GeV emission may not necessarily originate from a single component. For example, a joint model of an internal--external shock \citep{2011MNRAS.415...77M} hypothesizes that the external shock contribution alone cannot explain the initial bright spikes in the GeV band, and that a dominant internal shock contribution must be considered.
 Another example of the incompleteness of the synchrotron emission model is the observations of GRB 130427A \citep{2014Sci...343...42A}, which reveals that the synchrotron emission from the external shock in the late afterglow phase would not be enough to explain the observed  $\sim$ 50 GeV high-energy gamma-ray photons, which exceed the expected maximum synchrotron energy. One solution to explaining this violation could be to invoke synchrotron self-Compton (SSC) emission \citep{2017ApJ...844...92F}.

Thus, it is important to test the validity of the synchrotron model for an external shock to explain the observed behaviors in the GeV band from various views with a greater number of GRBs.

In this paper, gamma-ray emission observations of GRB 170405A made by the {\it Fermi}/LAT and GBM, which reveal delayed gamma-ray emission with respect to the X-ray emission, and a distinct feature of a high-energy spectral cutoff at $\sim$50 MeV that can significantly constrain the bulk Lorentz factor of the GRB ejecta, are reported.
In addition, optical follow-up observations by the UltraViolet Optical Telescope (UVOT) on {\it Swift} \citep{2004ApJ...611.1005G} and ground telescopes show that the early optical onset for this GRB occurred significantly later than the delayed onset in the GeV band followed by a fading afterglow, which implies that the gamma-ray emission did not originate through the external-shock model, and that other scenarios such as the internal-shock model should be explored to unveil the behavior in the GeV band.
This paper is organized as follows: the observations of {\it Fermi}, {\it Swift} and ground telescopes are described in Sect. \ref{sec:obs} and data analysis in Sect. \ref{sec:data_analysis}.
The obtained temporal and spectral characteristics are detailed in Sect. \ref{sec:lightcurve} and \ref{sec:spec}, respectively, and the physical interpretation is discussed in Sect. \ref{sec:discussion}. Finally, the conclusions of this study are summarized in Sect. \ref{sec:conclusion}.

\section{Observations}\label{sec:obs}
GRB 170405A triggered the {\it Fermi}/GBM and {\it Swift} Burst Alert Telescope (BAT) instruments simultaneously at 18:39:22.89 UT ($T_{\rm 0}$), 
and the X-Ray Telescope (XRT) and UVOT subsequently found a fading afterglow in the X-ray and optical bands, respectively, located at R.A., decl. = 14$\fh$39$\fm$18$\fs_.$72, -25$\fdg$14$\farcm$35$\farcs_.$3 (J2000) with a 90\% uncertainty of 0.61 arcsec \citep{2017GCN.20984....1T,2017GCN.20985....1E,2017GCN.20986....1H}. The {\it Fermi}/LAT also detected gamma-ray emission at a position consistent with the GRB, at 53$\arcdeg$ from the instrument boresight at the trigger time, and continued a follow-up observation through an autonomous repoint of the spacecraft. In addition, the LAT low energy (LLE) data in the 20--100 MeV range \citep{2010arXiv1002.2617P} display emission coincident with that in the $>$100 MeV band. The highest energy photon (810 MeV) associated with the GRB event was observed $\sim$450 s after the trigger.
After receiving alerts from the spacecraft, several ground telescopes promptly performed follow-up optical observations and detected the fading optical transient \citep{2017GCN.20988....1M,2017GCN.20989....1K,2017GCN.20991....1M,2017GCN.20992....1M,2017GCN.20996....1T,2017GCN.21003....1C}.
 In the $R$ band, we used values reported via the Gamma-ray Coordinates Network (GCN).
The redshift of this burst was determined to be {\it z} = 3.510 \citep{2017GCN.20990....1D}.

\section{Data Analysis}\label{sec:data_analysis}

For {\it Fermi}/GBM and LAT, to perform the temporal and spectral analyses, observational data was obtained from the {\it Fermi} Science Support Center\footnote{https://fermi.gsfc.nasa.gov/ssc/} and {\it Fermi} Science Tools (version v11r5p3)\footnote{https://fermi.gsfc.nasa.gov/ssc/data/analysis/software/} was used. When analyzing the LAT data within a short time interval ($\lesssim$100 s), events with the  ``{\tt P8R2\_TRANSIENT020E\_V6}" instrumental response function, which is a relatively loose-cut filter of the background events, was used, while the ``{\tt P8R2\_SOURCE\_V6}"-class events were used for the analysis over a longer time interval ($\gtrsim$100 s) to suppress background events. For the GBM data, four NaI detectors ({\tt n6}, {\tt n7}, {\tt n9}, and {\tt nb}) and one BGO detector ({\tt b1}) less than 50$^\circ$ from the instrument boresight at the trigger time were used.

For the {\it Swift} analysis, when performing the temporal and spectral analyses for BAT and XRT, the {\it Swift} Burst Analyser data\footnote{http://www.swift.ac.uk/burst\_analyser/00745797/} were utilized based on an automatic procedure provided by \cite{2009MNRAS.397.1177E}. For the UVOT analysis, {\tt uvotproduct} of the HEASoft package was used to perform photometry and generate the UV light curve. In the optical band, the optical magnitudes were corrected using dust maps, considering an extinction from the Milky Way \citep{1998ApJ...500..525S}\footnote{https://irsa.ipac.caltech.edu/applications/DUST/}, but an extinction from the host galaxy was not factored in as its extinction has a large uncertainty.

For the fitting procedure of the spectral analysis in the X-ray and gamma-ray bands, the {\it XSPEC} package \citep{1996ASPC..101...17A} was used.
In particular, to quantitatively estimate the goodness of the fit when performing the joint fitting of data, including LAT data, the ``$PG_{\rm stat}$" statistic, which is applicable to low-count statistics \citep{2011hxra.book.....A}, was used.

\section{Light curves}\label{sec:lightcurve}
\subsection{Prompt Emission}
\begin{figure*}[t]
\epsscale{1.6}
\plotone{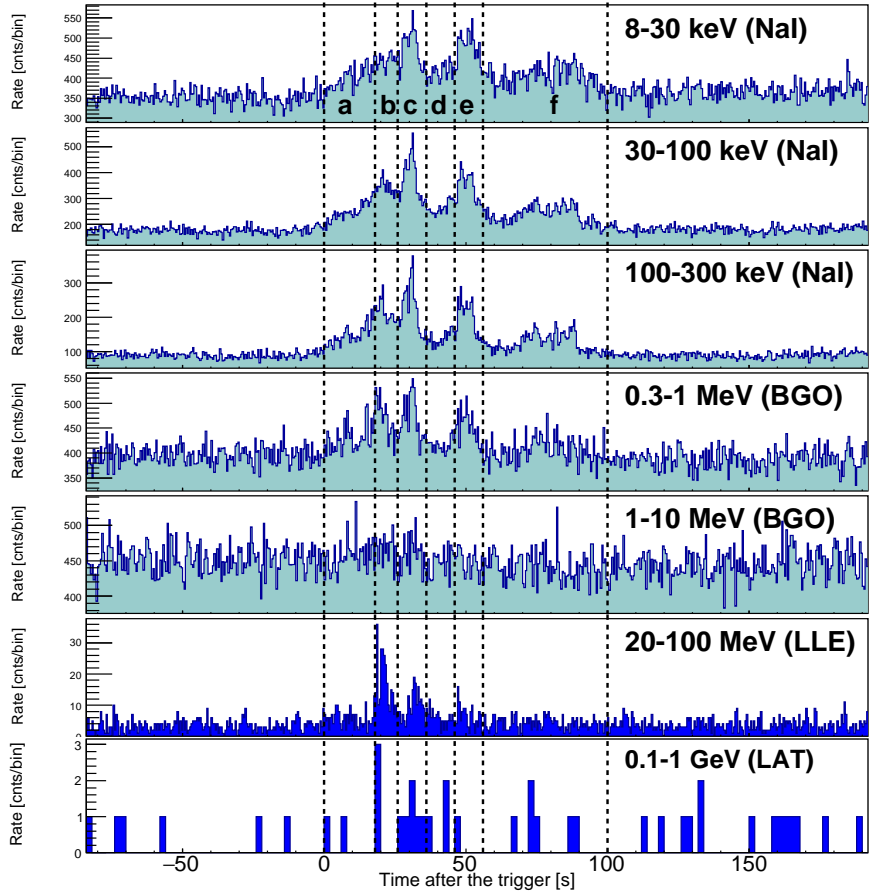}
\caption{Composite light curve of GRB 170405A. The top five panels are different bands in the GBM (NaI detectors for 8--30 keV, 30--100 keV and 100--300 keV and BGO detector for 0.3--1 MeV and 1--10 MeV), and the lower two are LAT. 
\label{compositeLC}}
\end{figure*}
Figure \ref{compositeLC} shows the composite light curves observed by the GBM and LAT detectors from 8 keV to 1 GeV.
  The duration of the GRB, expressed as $T_{90}$, is calculated to be 78.6 $\pm$ 0.6 s in the 50--300 keV band, where $T_{90}$ is the time interval between accumulating 5\% and 95\% of the total background-subtracted photons from the GRB event.
  From this, GRB 170405A is categorized as a long GRB. In the LLE band (20--100 MeV)  bright pulses were detected and gamma-ray emission above 20 MeV is delayed with respect to the X-ray emission below $\sim$1 MeV, which has been observed in other GRBs detected by the LAT \citep{2013ApJS..209...11A, 2FLGC}.
In the high-energy band above 1 GeV   there is no detection of gamma-ray photons by the LAT, which supports the hypothesis that in the prompt-emission phase, photons above 1 GeV are significantly damped.
The fluence of the GRB in the 10 keV --100 MeV range is calculated to be 1.55$^{+0.07}_{-0.06}$ $\times$ 10$^{-4}$ erg/cm$^2$ and the corresponding isotropic equivalent energy is $E_{\rm iso}$ = 4.03$^{+0.19}_{-0.17}$ $\times$ 10$^{54}$ ergs.

\subsection{Extended Emission} \label{sec:lcExtendedEmission}  
\begin{figure*}[t]
\epsscale{1.2}
\plotone{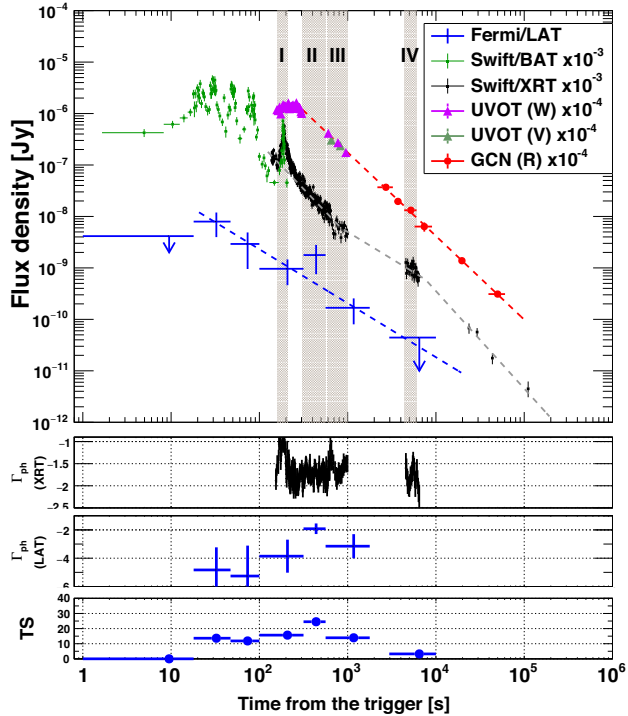}
\caption{Flux of the extended emission  detected by the  LAT (100 MeV--1 GeV) and afterglows detected  by the XRT  (0.3--10 keV) and UVOT onboard {\it Swift} and ground telescopes for GRB 170405A, where the optical fluxes are converted into the R-band flux (see the main text for details of the flux conversion). $\Gamma_{\rm ph}$ denotes the photon index. The photon index is fixed to $\Gamma_{\rm ph}$ = -2 when the upper limit of the LAT flux is calculated owing to low significance ({\it TS} $<$ 10). The dashed line represents the best-fit power-law function in the LAT data.  The  thick vertical  lines indicate the time interval of the joint SED discussed in Section \ref{sec:specExtendedEmission}.
} \label{lcExtendedEmission}
\end{figure*}
   Figure \ref{lcExtendedEmission} shows the temporally extended emission in the optical to gamma-ray bands.
   In the X-ray band, the light curves observed by the BAT and XRT show high temporal variability in the first 100 s, which is consistent with the {\it Fermi}/GBM observation.
   After that, the X-ray flux gradually decreases with time, while there is a bright X-ray flare at $\sim$200 s after the trigger.
   The X-ray temporal behavior, except for the X-ray flare, is fit well by a double-broken power-law function with temporal indices of $\alpha_{\rm AG}$ = -1.78$\pm$0.03 (100--840 s), -1.05$\pm$0.03 (840--7,600 s), and -1.90$\pm$0.29 (7,600--20,000 s).
   The X-ray photon indices measured by the XRT range from $\Gamma_{\rm ph}$ = -2.0 to -1.0, as illustrated in Fig. \ref{lcExtendedEmission}.

   The temporally extended emission in the gamma-ray band (100 MeV --1 GeV) was detected with a temporal index of $\alpha_{\rm AG}$ = -1.05$\pm$0.20 and {\it TS} $\sim$ 10--25, namely 3--5 $\sigma$, for each time bin, where {\it TS} is the test statistic to estimate the statistical significance \citep{1996ApJ...461..396M}.
   In the early phase, until $T_{\rm 0}$ + 100 s, the photon indices of the gamma-ray emission are much smaller than $\Gamma_{\rm ph}$ = -2, because the high-energy emission above $\sim$100 MeV is significantly damped, as will be discussed in Sect. \ref{sec:specPromptEmission}. In the later phase, although the uncertainty is large, the photon indices in the gamma-ray band gradually increase.
   At $T_{\rm 0}$ + 400 s, the flux rebrightening in the gamma-ray band leaves a signature, and correspondingly the photon index becomes harder, reaching $\Gamma_{\rm ph}$ $\sim$ -1.5.

   The UVOT detected the delayed onset in the $w$ band at $T_{\rm 0}$ + $\sim$200 s, after which the optical flux decreased as a single power-law function with a temporal index of $\alpha_{\rm AG}$ = -1.54$\pm$0.08. Furthermore, the temporal indices in the $R$ and $v$ bands are consistent with that in the $w$ band, as shown in Table \ref{table:alpha_EE}.
   It should be noted that in the case of GRB 170405A with {\it z} = 3.51, emission above 1216 $\AA$ $\times (1+z)$ $\sim$ 5480 $\AA$ would be significantly damped due to hydrogen absorption, which moderately affects the estimate of intrinsic flux in the {\it w} and {\it v} bands, while the emission in the $R$ band is not affected by any significant margin.
 Flux correction of the extinction by hydrogen absorption is difficult in the $w$ band because the bandwidth is exceedingly wide.
Here, because the temporal decay indices in the {\it w}, {\it v},  and {\it R} bands are consistent in the 1-$\sigma$ confidence range, as shown in Table \ref{table:alpha_EE} (although the uncertainty of the index in the {\it v} band  is very large), it would be reasonable to claim that the emissions in the {\it w}, {\it v}, and {\it R} bands have the same origin. When plotting the light curves in the {\it w}, {\it v}, and {\it R} bands shown in Fig. \ref{lcExtendedEmission}, the fluxes in the {\it w} and {\it v} bands were converted into those in the $R$ band by multiplying with 17.0 and 2.8, respectively, to let those fluxes have the same baseline, so that the temporal behavior in the optical band could be easily visualized.
 
 \begin{table*}[hb]
\begin{center}
\caption{Fitting result of the temporal indices during the extended emission\label{table:alpha_EE}}
\footnotesize 
\begin{tabular}{ccccccc}
\tableline\tableline

filter & $\alpha_{\rm AG1}$  & $t_{\rm 1}$ [s] &  $\alpha_{\rm AG2}$ & $t_{\rm 2}$ [s] & $\alpha_{\rm AG3}$ & $\chi^2$/d.o.f.  
\\
\tableline
LAT & -1.05$\pm$0.20 & -- & -- & -- & -- &  1.7/3 \\
$R$  & -1.63$\pm$0.03 &  -- & -- & -- & --   &  1.2/4   \\
$w$  & -1.54$\pm$0.08 &   -- & -- & -- & --   & 4.2/11   \\
$v$  & -1.07$\pm$1.04 &   -- & -- & -- & --    & 4.6$\times$10$^{-7}$/0   \\
XRT & -1.78$\pm$0.03 &  (8.41$\pm$0.04)$\times$10$^{2}$  & -1.05$\pm$0.03 & (7.60$\pm$4.25)$\times$10$^{3}$ &  -1.89$\pm$0.29 & 229.2/161 \\
\tableline
\end{tabular}
\tablecomments{In the X-ray band, the fitting function consists of a double-broken power-law function with three photon indices ($\alpha_{\rm AG1}$, $\alpha_{\rm AG2}$ and $\alpha_{\rm AG3}$) and two break times ($t_{\rm 1}$ and $t_{\rm 2}$) during the afterglow (AG) phase. Here, the time interval from 160 s to 300 s, in which a bright X-ray flare occurred, was excluded from the fitting. \\
In the $w$ band, because the emission in the time interval from $T_{\rm 0}$ + 150 s to $T_{\rm 0}$ + 250 s is almost constant, a power-law function was fit in the time interval from 250 s. \\ Errors correspond to the 1-$\sigma$ confidence region.
}
\end{center}
\end{table*}

\section{Spectrum}\label{sec:spec}
\subsection{Prompt emission}\label{sec:specPromptEmission}  
\begin{figure*}[th]
\epsscale{1.2}
\plotone{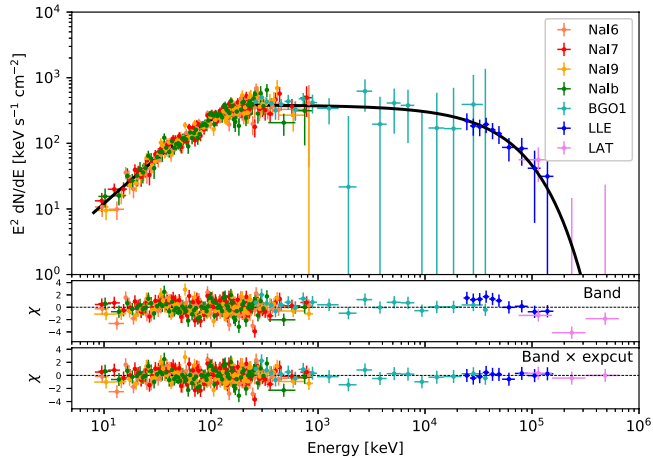}
\caption{Observed spectrum in the prompt-emission phase (Interval $b$: $T_0$ +18 s to + 26 s) detected with GBM and LAT. $Top$: The best-fitting function (i.e., the Band function with an exponential cutoff). $Middle$: Residuals from the Band function {\it without} an exponential cutoff. $Bottom$: Residuals from the Band function {\it with} an exponential cutoff.
 \label{spec:cutoff}}
\end{figure*}
In the prompt emission phase, the whole interval was divided into six parts to characterize the time-resolved spectral features.
In each time interval, the Band function \citep{1993ApJ...413..281B} could be applicable as shown in Table \ref{table:spec_prompt}.
However, in some time intervals, statistical improvements were observed whenever an exponential cutoff was added to the Band function in the high-energy band.
In the brightest part of the time interval  ``$b$" where delayed sub-GeV emission was detected, a wavy structure of the residual can be seen around $\sim$50 MeV, as shown in Fig. \ref{spec:cutoff}. However, if an exponential cutoff is added to the Band function, the wavy residual structure is significantly reduced,  which suggests that there is most likely a spectral cutoff in the high-energy range.

To quantitatively estimate the statistical evaluation of the improvement (i.e., goodness of fit) by adding the spectral cutoff, the differences between PG$_{\rm stat}$ values of the Band functions with/without an exponential cutoff, denoted as $\Delta$PG$_{\rm stat}$, were calculated to be 45 for an increase of one degree of freedom.
To estimate the statistical significance of the exponential cutoff, a Monte-Carlo method was used with the XSPEC package. First, one million simulated spectra were generated from the fitting result of the Band function without an exponential cutoff in the interval  ``$b$". The generated spectra were then fit with the Band functions, with and without an exponential cutoff, which correspond to the null hypothesis and alternative models, respectively. For each generated spectrum, $\Delta$PG$_{\rm stat}$ between the null hypothesis and alternative models was calculated \citep[similar procedures for the statistical evaluation in detail are also reported at][]{2013ApJS..209...11A, 2016ApJ...833..139A}.
From the simulated dataset of one million spectra, the largest $\Delta$PG$_{\rm stat}$ obtained was 37, which means a null hypothesis probability of $<$ 10$^{-6}$ that the best-fitting function is the Band function with an exponential cutoff, if the true function is the Band function without an exponential cutoff.

\begin{table*}[h]
\begin{center}
\caption{Spectral fitting results from the LAT and GBM in the prompt-emission phase.\label{table:spec_prompt}}
\footnotesize 
\begin{tabular}{clccccc}
\tableline\tableline
Interval & Model & $\alpha$ & $\beta$ & $E_{\rm peak}$ & $E_{\rm cut}$ &  PGstat/dof 
\\
{[s]}  & & &  & [keV] & [MeV] &     \\
\tableline
``total"  & Band & $-$0.77$\pm$0.03 & $-$2.43$\pm$0.03  & 277$\pm$11 & --- &   2725.2/608   \\
 (0--100 s) &  Band $\times$ expcut & $-$0.74$\pm$0.03 & $-$2.21$^{+0.05}_{-0.06}$  & 259$\pm$12 & 68$^{+26}_{-16}$ &   2668.6/607 \\
   \hline
   
``$a$" & Band & $-$0.68$^{+0.08}_{-0.07}$ & $-$2.51$^{+0.08}_{-0.10}$  & 340$^{+43}_{-35}$ & --- &   1003.0/608   \\
 (0--18 s) & Band $\times$ expcut & $-$0.63$^{+0.09}_{-0.08}$ & $-$2.17$\pm$0.19  & 311$^{+44}_{-40}$ & 42$^{+64}_{-20}$ &  992.0/607 \\
\hline

``$b$" & Band & $-$0.56$\pm$0.08 & $-$2.27$\pm$0.04 & 293$^{+25}_{-22}$ & --- &  806.2/609   \\
 (18--26 s) & Band $\times$ expcut & $-$0.50$\pm$0.08 & $-$2.01$\pm$0.07  & 262$^{+27}_{-25}$ & 48$^{+23}_{-13}$ &  761.2/608 \\
\hline

``$c$"  & Band & $-$0.73$\pm$0.05 & $-$2.43$\pm$0.11 & 303$^{+22}_{-20}$ & --- &   829.2/608   \\
(26--36 s) & Band $\times$ expcut & $-$0.71$\pm$0.08 & $-$2.20$\pm$0.07  & 285$^{+24}_{-23}$ & 66$^{+68}_{-26}$ &   814.1/607 \\
\hline

``$d$" & Band & $-$0.63$^{+0.14}_{-0.12}$ & $-$2.30$^{+0.06}_{-0.08}$ & 193$^{+29}_{-24}$ & --- &   843.9/608   \\
 (36--46 s) & Band $\times$ expcut & $-$0.56$^{+0.16}_{-0.15}$ & $-$2.11$^{+0.12}_{-0.14}$  & 173$^{+32}_{-25}$ & 84$^{+218}_{-40}$ &   837.5/607 \\
\hline

``$e$" (46--56 s)  & Band & $-$0.86$\pm$0.05 & $-$2.59$^{+0.14}_{-0.19}$ & 284$^{+27}_{-23}$ & --- &  794.0/598   \\
\hline

``$f$" (56--100 s)  & Band & $-$0.87$\pm$0.06 & $-$2.55$^{+0.10}_{-0.13}$ & 232$^{+23}_{-20}$ & --- &  1606.7.9/598   \\
\hline
\tableline
\end{tabular}
\tablecomments{Band function parameters are low-energy photon index $\alpha$, high-energy photon index $\beta$, and peak energy $E_{\rm peak}$ with an exponential cutoff (``expcut") at the cutoff energy $E_{\rm cut}$. 
Errors correspond to the 90\% confidence region. In the intervals of  ``$e$" and  ``$d$", when applying the  ``Band $\times$ expcut" function, no meaningful limit on the cutoff energy could be obtained, and only the result of the Band function could be shown.
}
\end{center}
\end{table*}

\subsection{Extended emission}\label{sec:specExtendedEmission}  
\begin{figure*}[ht]
  \begin{center}
   \begin{minipage}[xbt]{50mm}
    \resizebox{50mm}{!}
    {\includegraphics[angle=0]{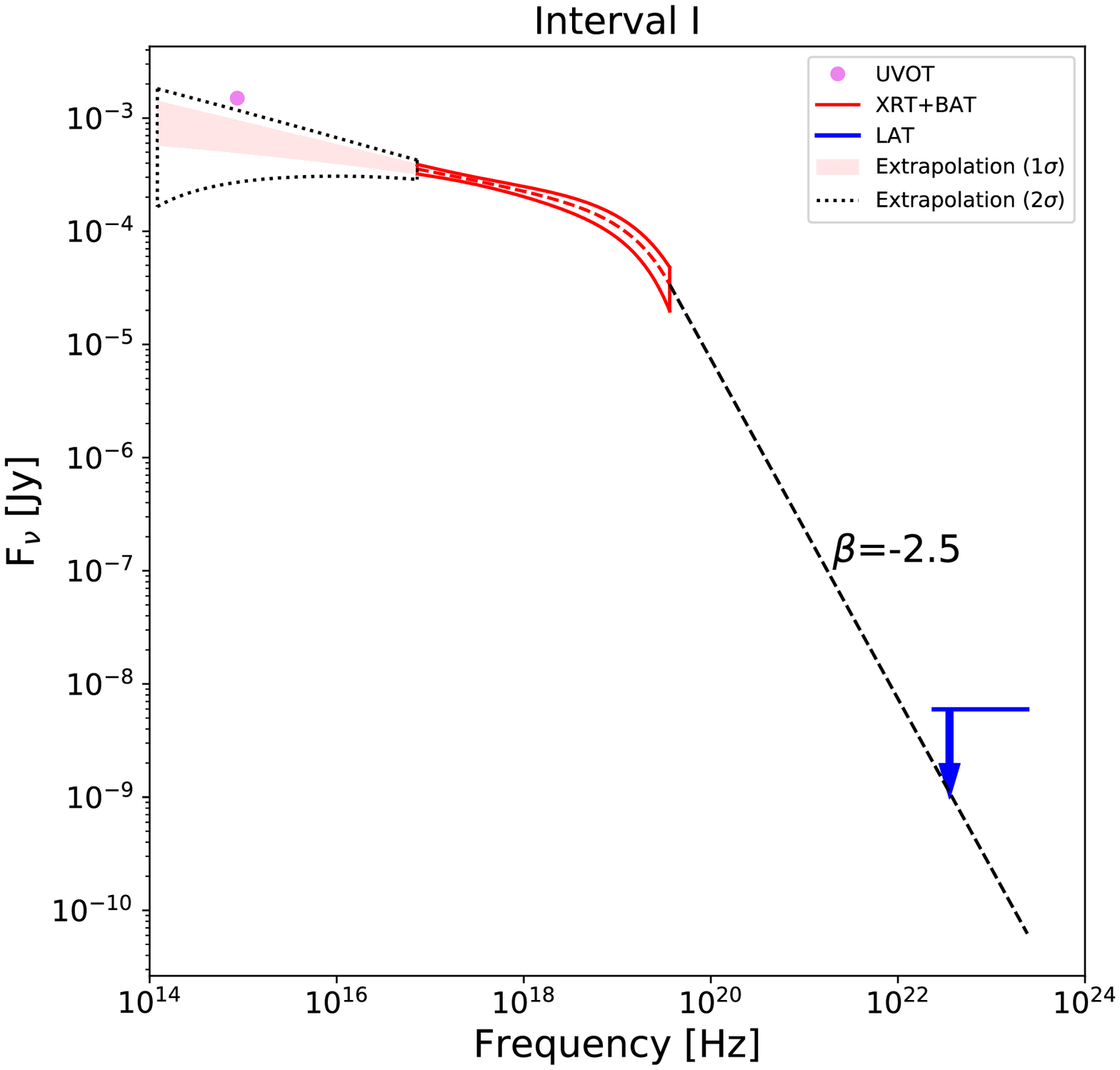}}
   \end{minipage}
   \begin{minipage}[xbt]{50mm}
    \resizebox{50mm}{!}
    {\includegraphics[angle=0]{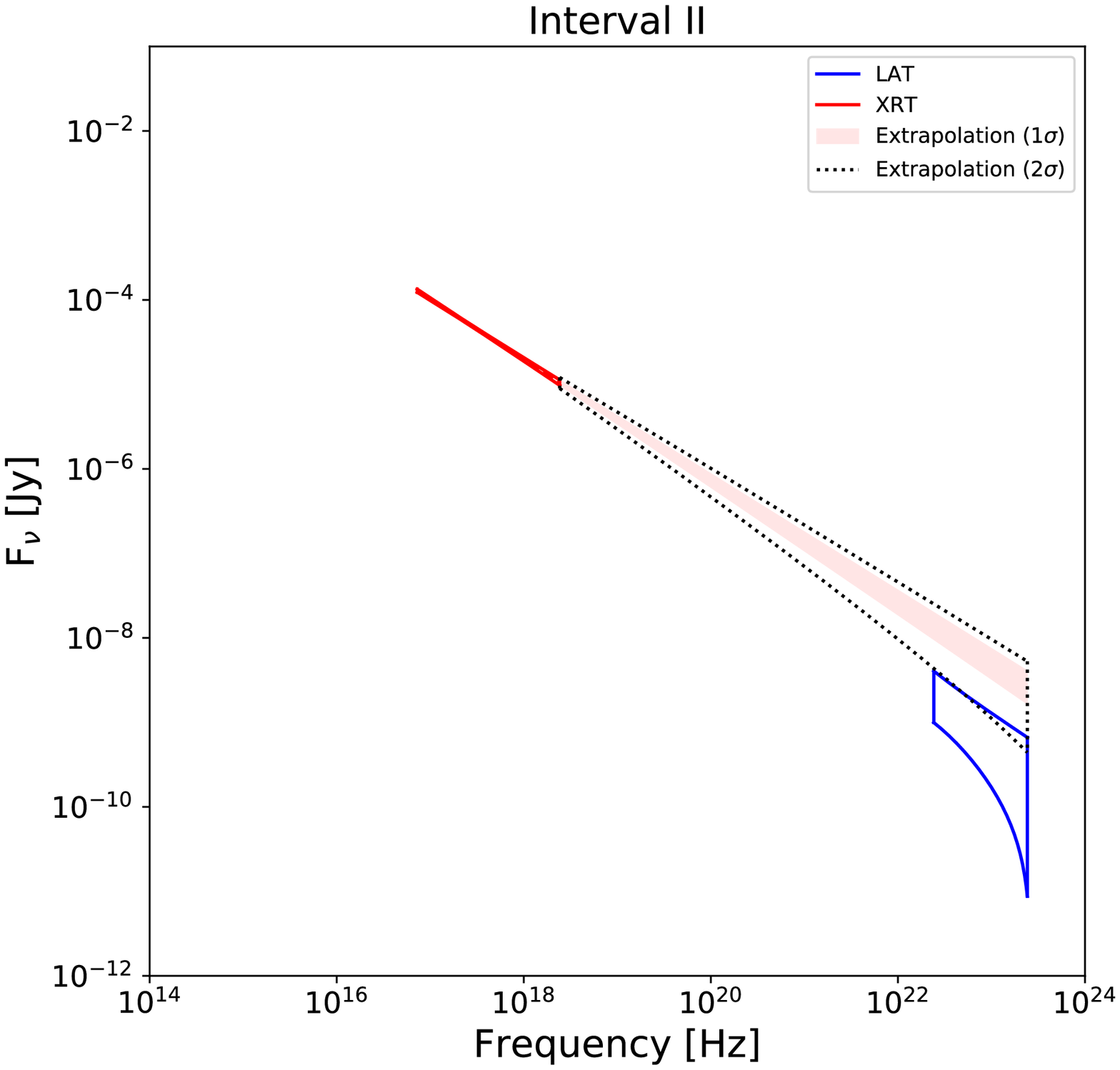}}
   \end{minipage}
    \end{center}
  \begin{center}
   \begin{minipage}[xbt]{50mm}
    \resizebox{50mm}{!}
    {\includegraphics[angle=0]{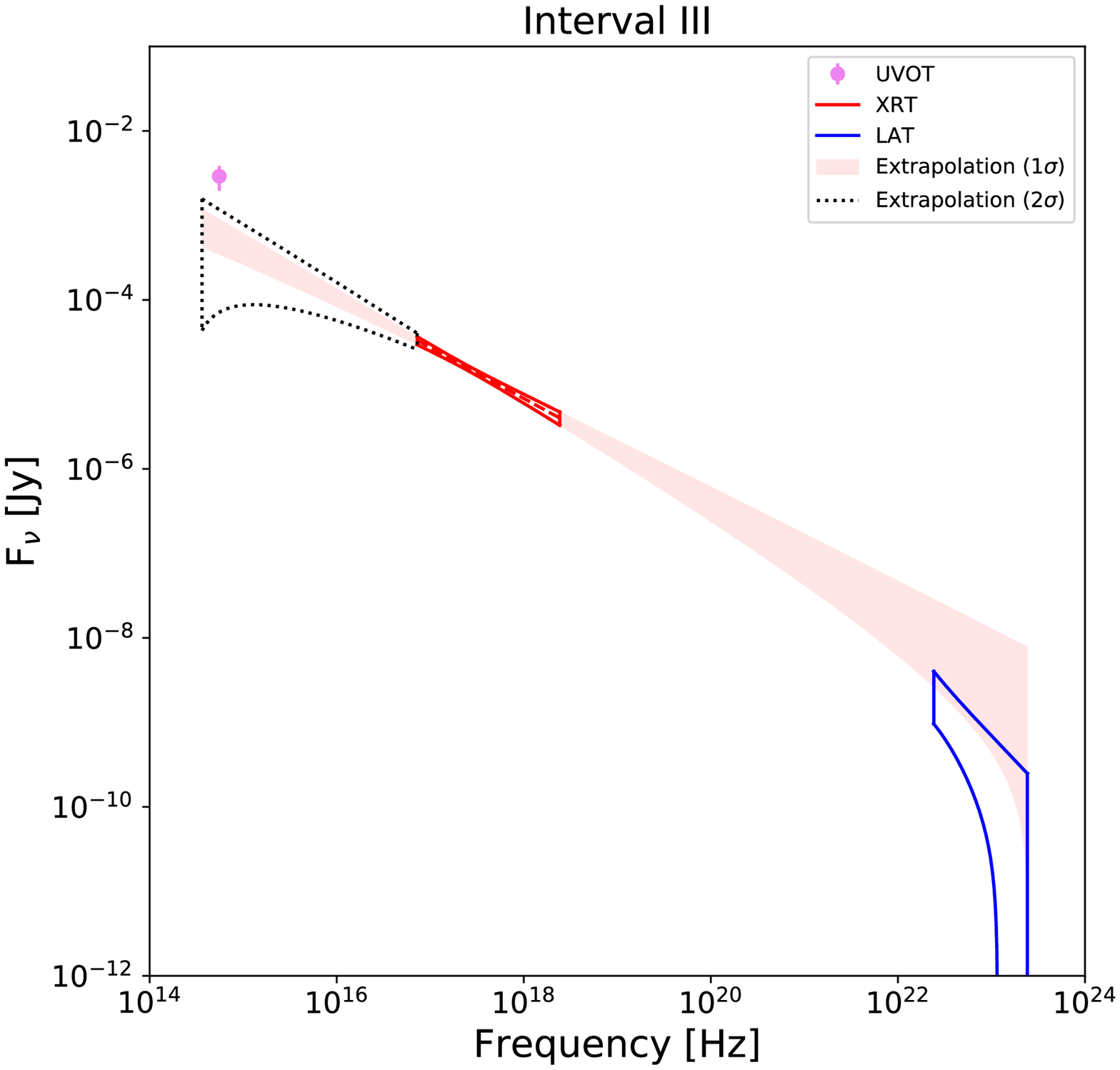}}
   \end{minipage}
   \begin{minipage}[xbt]{50mm}
    \resizebox{50mm}{!}
    {\includegraphics[angle=0]{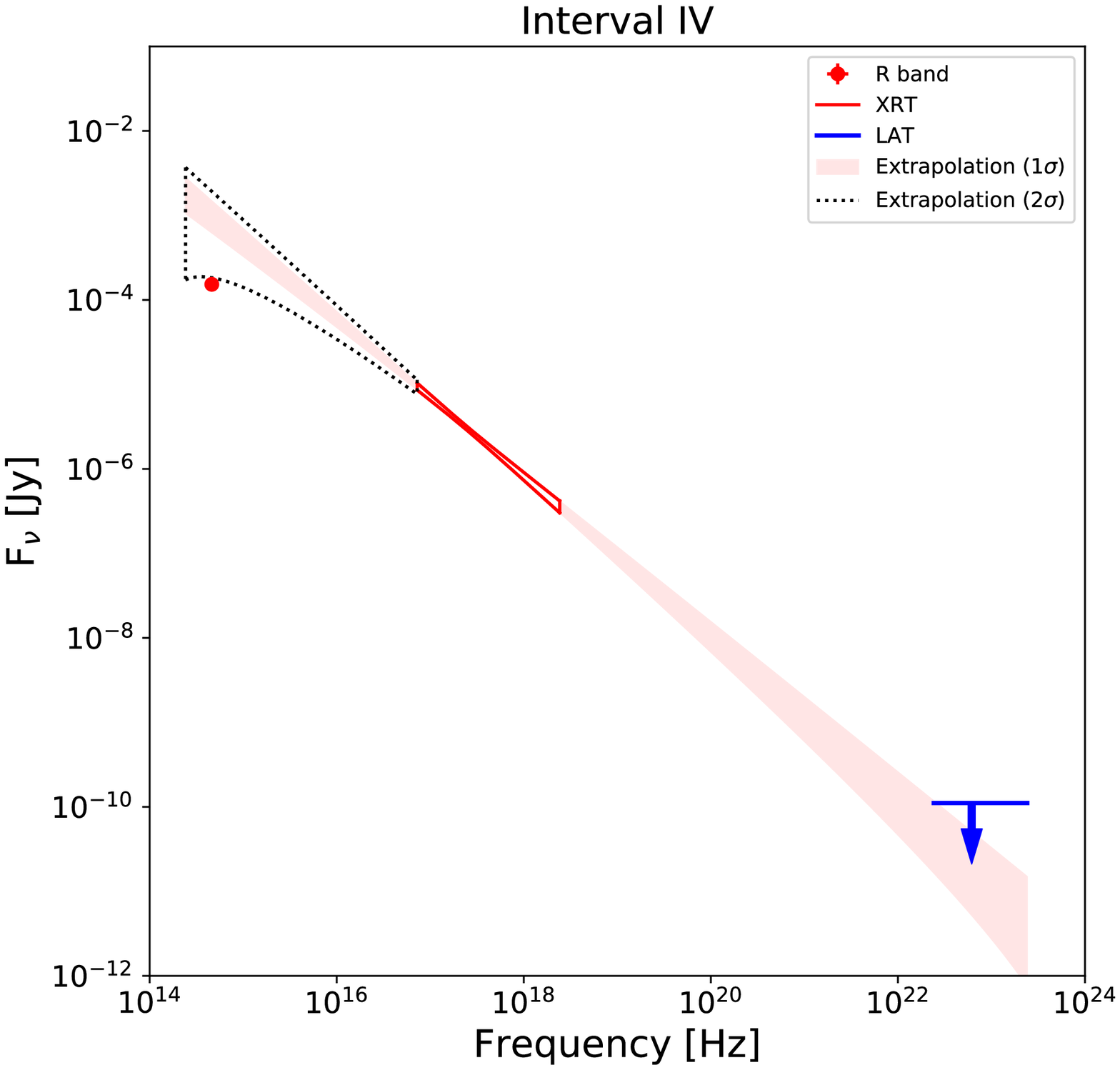}}
   \end{minipage}
   \end{center}
     \begin{center}
   \caption{Spectral energy distributions from the optical to gamma-ray bands for four time intervals: Intervals I, II, III, and IV denote the shaded regions shown in Fig. \ref{lcExtendedEmission}. The solid red and blue areas correspond to the 1-$\sigma$ region for the XRT + BAT and LAT ranges, respectively. The pale red shaded and dotted regions correspond to the 1-$\sigma$ and 2-$\sigma$ regions extrapolated from the XRT band. The blue arrows represent the 95\% upper limit for the LAT. For interval I, the extrapolation above the BAT range is shown assuming that the high-energy photon index is -2.5.  }
   \label{specExtendedEmission}
  \end{center}
 \end{figure*}

The spectral energy distributions (SEDs) from the four time intervals  ``I",   ``I\hspace{-.1em}I",  ``I\hspace{-.1em}I\hspace{-.1em}I" and  ``I\hspace{-.1em}V" in Fig. \ref{lcExtendedEmission} are shown in Fig. \ref{specExtendedEmission}, and the fitting results in the X-ray band are given in Table \ref{table:spec_extend}. Note that every optical/UV flux is corrected considering only the Galactic extinction, and the correction of extinction from the host galaxy is not included.
 When fitting the X-ray spectra, a fixed Galactic absorption\footnote{http://www.swift.ac.uk/xrt\_spectra/00745797/} of 8.75$\times$10$^{20}$ cm$^{-2}$ was used.
 As illustrated in Table \ref{table:spec_extend}, the absorption from the host galaxy is negligible in all time intervals.

In the time interval  ``I" where rebrightening in the X-ray band (or X-ray flare) occurred, the obtained X-ray spectrum is well represented by the cutoff power-law function.
The UV flux is located beyond an extrapolation from the X-ray spectrum with more than 2-$\sigma$ confidence level, which indicates that the X-ray and optical emissions have different origins. 

In the time interval  ``I\hspace{-.1em}I", where no optical data is available and a flux increment in the GeV band was observed, the SED shows that the power-law function between the X-ray and GeV bands is almost consistent. 

In the time intervals  ``I\hspace{-.1em}I\hspace{-.1em}I" and  ``I\hspace{-.1em}V",  there are no significant temporal structures in any of the energy bands.
 In time interval  ``I\hspace{-.1em}I\hspace{-.1em}I" the optical flux is located beyond an extrapolation from the X-ray spectrum, similar to the trend in interval  ``I". Contrary to interval ``I\hspace{-.1em}I\hspace{-.1em}I", in time interval  ``I\hspace{-.1em}V" the optical flux is far below extrapolation from the X-ray spectrum with a 2-$\sigma$ confidence level.
This feature can be explained by the fact that the temporal flux decrease in the optical/UV band is steeper than that in the X-ray band, as shown in Fig. \ref{lcExtendedEmission} and Table \ref{table:alpha_EE}. 

In each time interval, because the GeV emission has a large degree of uncertainty or an upper limit, there are no strong constraints in regard to the SEDs. However, the optical/UV emissions are not consistent with the extrapolation from the X-ray emission, which indicates separate origins, and such different behaviors are also confirmed by the observation of the temporal evolution between the optical and X-ray bands as shown in Sect. \ref{sec:lcExtendedEmission}.

\begin{table*}[h]
\begin{center}
\caption{Spectral fitting results  in the extended-emission phase.\label{table:spec_extend}}
\footnotesize 
\begin{tabular}{clcccccc}
\tableline\tableline

Interval & Instrument & Model & $N_{\rm H, gal}$ & $N_{\rm H, host}$ & Photon index & $E_{\rm peak}$ &  $\chi^2$/dof 
\\
{[s]}  & & &10$^{20}$ cm$^{-2}$  & 10$^{20}$ cm$^{-2}$  & & [keV] &     \\
\tableline
``I"  &  XRT+BAT &  Cutoff PL & 8.75 (fixed) & 1.07$^{+unc}_{-unc}$$\times$10$^{-6}$ &  $-$1.19$\pm$0.06  & 103$^{+76}_{-30}$  &   499.6/589   \\
 (180--200 s) &  LAT  & PL & ---  & ---  & --- & --- &   --- \\
   \hline
  ``I\hspace{-.1em}I"  &  XRT &  PL &  8.75 (fixed) & 3.37$^{+unc}_{-unc}$$\times$10$^{-6}$ &  $-$1.79$\pm$0.05  & ---  &   516.9/565   \\
 (310--560 s) &  LAT  & PL & --- & ---  &--1.88$\pm$0.33$^\dagger$   & --- &   --- \\
   \hline 
  ``I\hspace{-.1em}I\hspace{-.1em}I"  &  XRT &  PL & 8.75 (fixed) &  4.30$^{+unc}_{-unc}$$\times$10$^{-7}$ &  $-$1.68$\pm$0.12  & ---  &   214.8/243   \\
 (589--1000 s) &  LAT  & PL & --- & ---  &--2.36$\pm$0.50$^\dagger$ & --- &   --- \\
   \hline 
  ``I\hspace{-.1em}V"  &  XRT &  PL & 8.75 (fixed) & 8.78$^{+unc}_{-unc}$$\times$10$^{-7}$ &  $-$2.01$\pm$0.11  & ---  &   236.1/263   \\
 (4375--6000 s) &  LAT  & PL & --- & ---  &--- & --- &   --- \\
   \hline 

\tableline
\end{tabular}
\tablecomments{Fitting function in the X-ray band consists of a single power-law function with the Galactic and host-galaxy absorptions (and a spectral cutoff when using the BAT data), namely {\tt TBabs $\times$ zTBabs $\times$ PL (or Cutoff-PL)}.
 The power-law function parameters are the low-energy photon index $\alpha$, the high-energy photon index $\beta$, and the peak energy $E_{\rm peak}$ with an exponential cutoff (``expcut") at the cutoff energy, $E_{\rm cut}$. 
Errors correspond to the 90\% confidence region. ``$unc$" means that the fitting parameter is unconstrained.\\
$^\dagger$: Errors correspond to the 1-$\sigma$ confidence region.
}
\end{center}
\end{table*}

\section{Discussion}\label{sec:discussion}

\subsection{Bulk Lorentz factor}\label{sec:BulkLorentzFactor}
A significant spectral cutoff was detected during the prompt emission, which could be naturally interpreted as pair-production opacity within the GRB source \citep{2011ApJ...729..114A, 2015ApJ...806..194T, 2018ApJ...864..163V}.
If a certain emission volume is considered, high-energy gamma-ray photons would interact with low-energy photons via pair creation.
The opacity of gamma-ray photons as a function of the photon energy and the photon density in an emission region is controlled by the bulk Lorentz factor, $\Gamma_{\rm bulk}$.
 In such a case, this spectral cutoff feature can be used as an estimator of the bulk Lorentz factor of the GRB ejecta.
Here, the dominant emission component comes from the Band component, and its high-energy segment, which contributes to the annihilation of gamma-ray photons, can be accurately represented as a power-law function. 
 A detailed numerical calculation considering the temporal, spatial and directional dependence for the pair-production interaction \citep[][]{2008ApJ...677...92G} gives  a conservative lower limit of the bulk Lorentz factor, $\Gamma_{\rm bulk,min}$, which likely corresponds to a realistic case. Here we use Equation (1) of \cite{2019arXiv190910605A} with observed parameters of the photon spectrum with a spectral cutoff of $E_{\rm cut}$ and  the variability timescale $\Delta t$ (i.e., the emission radius is $R$ = $\Gamma_{\rm bulk}^2 c \Delta t /(1+z)$). 



\begin{figure}[t]
\epsscale{1.0}
\plotone{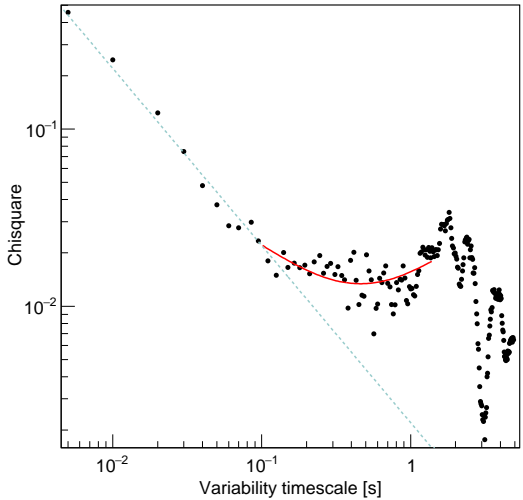}
\caption{Power spectrum of the LLE light curve for GRB 170405A from $T_{\rm 0}$ + 15 s to 40 s. The circles represent the calculated spectrum from the observed data. The dashed line represents the power $\propto$ $\Delta t^{-1}$, meaning Poisson statistical fluctuation. The red solid line represents the best-fit curve when using the log-parabola function.
 \label{powspec:min_var_time}}
\end{figure}

The observed variability timescale, $\Delta t$, can be estimated from a temporal analysis that calculates the power of a differentiated time-series of a light curve, by subtracting photon counts in an individual time bin from those in the adjacent bin \citep{1997JGR...102.9659N}. By using the LLE lightcurve from $T_{\rm 0}$ + 15 s to 40 s, the obtained power spectrum is shown in Fig. \ref{powspec:min_var_time}. When estimating the variability timescale from the calculated spectrum, the Poisson noise in the light curve should be considered, in which the noise is represented as the power, $\propto$ $\Delta t^{-1}$.
 If an additional component can be found above the noise, the component would correspond to the variability timescale.
Then, the variability timescale is defined as the minimum point of power when representing through the log-parabola function, where the minimum point corresponds to an apparent turnaround from the background to the GRB emission.
  The variability timescale parameter for GRB 170405A is found to be 0.46 $\pm$ 0.04 s. The variability timescale was also examined by calculating the pulse width of the bright narrow spike in the LLE light curve at $T_{\rm 0}$ + 18.5 s by fitting it with a Gaussian function, and its width was obtained as 0.53 $\pm$ 0.08 s (full width at half-maximum: FWHM), which is consistent with the result obtained from the power-spectrum method.

Using the measured variability timescale of $\Delta t$ $\sim$ 0.5 s, $E_{\rm cut}$ = 48 MeV and the observed photon spectrum, the bulk Lorentz factor was obtained as   $\Gamma_{\rm bulk, min}$ = 170. On the other hand, if the cutoff energy in the jet-comoving frame is  $m_e c^2$, which corresponds to a maximum Lorentz factor \citep{2018MNRAS.475L...1G}, $\Gamma_{\rm bulk, max}$ =  $(1+z) E_{\rm cut}/ m_e c^2$ = 420. From these considerations, the bulk Lorentz factor for this burst is estimated to be $\Gamma_{\rm bulk}$ = 170 -- 420.

\subsection{Rebrightening in the X-ray band}
In the X-ray band, the intense rebrightening at $T_{\rm 0}$ + 200 s occurred simultaneously with the optical onset, which may indicate that the X-ray emission has the same origin as the optical emission. However, as shown in the obtained SED result at Interval ``I" (Fig. \ref{specExtendedEmission}), the optical flux significantly exceeds the extrapolation from the X-ray emission, and it is not evident if the X-ray and optical emissions have the same origin. In addition, the obtained photon index in the X-ray band is $\Gamma_{\rm ph}$ = -1.2, which is harder than that of typical afterglows \citep{2009ApJ...698...43R}. Furthermore, such a hardening feature in the X-ray band during the rebrightening can be seen in Fig. \ref{lcExtendedEmission} (see the second top panel: time vs. $\Gamma_{\rm ph}$ in the X-ray band with XRT). In addition, the temporal indices of the X-ray rebrightening are $\alpha_{\rm AG}$ = +7.08$\pm0.07$ and -6.12$\pm$0.16 in the rise and decay parts, respectively, which are far from the typical photon indices expected from an external-shock origin: $t^{+2}$ or even t$^{+11/3}$ in the rise part and t$^{\sim-1.2}$ if the electron index is $p$ $\sim$ 2.3 in the decay part \citep{1999ApJ...520..641S}.
In regard to these observational results, it is reasonable to conclude that the X-ray rebrightening at $T_{\rm 0}$ + 200 s has a different origin, such as a late internal shock, and is not related with the external shock.

\subsection{Optical and GeV onsets}\label{sec:optical_GeV_onset}
Interestingly, the multi-wavelength observations reveal different temporal onsets in the optical and GeV bands at $T_{\rm 0}$ + $\sim$200 s and $\sim$20 s, respectively.
 Such delayed temporal onsets can be interpreted as the beginning of the external forward shock \citep{1999ApJ...520..641S}. However, if both onsets have the same physical origin and correspond to the deceleration time, they should occur at the same time. Thus, the different observed temporal onsets indicate that the GeV and optical emissions have different physical origins.
In the optical band, after the optical onset, the optical flux monotonically decreases as a single power-law function ($\propto t^{-1.6}$)  until $\sim$10$^5$ s.
 This result implies that the optical emission should be interpreted to be of external-shock origin, and the GeV emission with an earlier onset than that in the optical band to be of a different origin. In addition, the spectral evolution in the GeV band would support a non-external shock origin. Hereafter, we investigate whether this interpretation is valid by discussing the external-shock origin to explain the GeV emission in detail:

 First, let us assume $\Gamma_{\rm bulk}$ = 170 derived in Sect. \ref{sec:BulkLorentzFactor}. That can be used to estimate the afterglow onset in the ambient profile of the interstellar medium (ISM) \citep{1999ApJ...520..641S}, which is formulated as
\begin{equation}
\begin{split}
t_{\rm onset, ISM} &= \biggl[   \frac{3E_{\rm iso} (1+z)^3}{32 \pi n_{\rm ISM}m_{\rm p}c^5  \eta \Gamma_{\rm bulk}^8 } \biggr]^{1/3} \\ &\sim  200\,  {\rm s} \left( \frac{E_{\rm iso}}{4 \times 10^{54}\, {\rm ergs} }\right)^{1/3}\left( \frac{\eta}{0.1} \right)^{-1/3} \\
& \quad\quad\left( \frac{n_{\rm ISM}}{ 500\,{\rm cm}^{-3}} \right)^{-1/3} \left( \frac{\Gamma_{\rm bulk}}{ 170} \right)^{-8/3} \nonumber
\end{split}
\label{eq:t_onset_ISM}
\end{equation}
where $n_{\rm ISM}$ is the density of the ISM, $m_{\rm p}$ is the proton mass, and $\eta$ is the efficiency of total shock energy in converting into the gamma-ray emission. 
 If $\eta$ = 0.1 and $n_{\rm ISM}$ = 500 cm$^{-3}$, then the expected afterglow onset is consistent with the observed optical onset ($T_{\rm 0}$ + $\sim$200 s). However, the very high density required is not realistic. Furthermore, to explain the gamma-ray onset observed at $T_{\rm 0}$ + $\sim$20 s in the context of external shock, a higher ISM density of $\sim$5 $\times$ 10$^{5}$ cm$^{-3}$ is needed. Thus, the ISM scenario is unlikely if $\Gamma_{\rm bulk, min}$ = 170.  Instead, if we adopt $\Gamma_{\rm bulk, max}$ = 420 and  $n_{\rm ISM}$ = 0.3 cm$^{-3}$, the observed optical onset at $T_{\rm 0}$ + $\sim$200 s can be explained by an afterglow origin.

In addition, this study also considers the afterglow in a stellar wind environment \citep{2000ApJ...536..195C}. In that case, the afterglow onset can be estimated as 
\begin{equation}
\begin{split}
t_{\rm onset, wind} &=    \frac{E_{\rm iso} (1+z)}{16 \pi A m_{\rm p}c^3  \eta \Gamma_{\rm bulk}^4 }  \\ 
&\sim  200\,  {\rm s} \left( \frac{E_{\rm iso}}{4 \times 10^{54}\, {\rm ergs} }\right)\left( \frac{\eta}{0.1} \right)^{-1} \\
&\quad\quad \left( \frac{A^{*}}{ 1.5} \right)^{-1} 
\left( \frac{\Gamma_{\rm bulk}}{ 170} \right)^{-4} \nonumber
\end{split}
\label{eq:t_onset_wind}
\end{equation}
 where the wind parameter, $A$ = 3.02 $\times$ 10 $^{35}$ $A^*$ cm$^{-1}$, when the mass-loss rate is 1 $\times$ 10$^{-5}$ $M_{\sun}$ yr$^{-1}$ and the wind velocity is 10$^{3}$ km s$^{-1}$ for $A^*$ = 1. If we take $\eta$ = 0.1 and $A^*$ =  1.5, then the optical onset at $T_{\rm 0}$ + 200 s can be interpreted as being of the external-shock origin. However, the expected onset time is sensitive to $A^*$ and can easily change depending on $A^*$. For example, if $A^*$ =  20, then $t_{\rm onset, wind}$ $\sim$  20 s, which is consistent with the gamma-ray onset time. Thus, the external-shock scenario is tested by confirming the consistency between the observed and expected fluxes in detail. Hereinafter, we discuss the external-shock scenario from the perspective of both ISM and wind cases.

\subsection{Afterglow scenario}\label{subSec:AG_scenario}
\subsubsection{ISM case}
As mentioned in the previous section, in the case of ISM,  if we use $\Gamma_{\rm bulk, max}$ = 420, the optical onset can be interpreted as the early epoch of the external-shock emission.  Here, using the GeV onset at $T_{\rm 0}$ + 20 s as a deceleration timescale requires $n_{\rm ISM}$ $\sim$ 200 cm$^{-3}$, which is a very high ISM density.
 In such a case, there are two plausible explanations for the optical emission, which are (1) $\nu_{\rm m} < \nu_{\rm opt} < \nu_{\rm c}$ and (2) $\nu_{\rm m} <  \nu_{\rm c} < \nu_{\rm opt}$ (or $\nu_{\rm c} <  \nu_{\rm m} < \nu_{\rm opt}$), where $\nu_{\rm m}$ is the frequency of the synchrotron emission from electrons with a minimum Lorentz factor, $\nu_{\rm c}$ is the cooling frequency of the synchrotron emission, and $\nu_{\rm opt}$ is a frequency in the optical band ($\sim$ 10$^{15}$ Hz).  We can describe $\nu_{\rm m}$, $\nu_{\rm c}$,  and the crossing time $t_{\rm c}$ when $ \nu_{\rm c}$ crosses the observed frequency $ \nu_{\rm obs}$ as follows  \citep{1998ApJ...497L..17S}, 
\begin{equation}
\begin{split}
\nu_{\rm m} &\sim 3 \times 10^{21} \epsilon_{\rm e}^{2}\epsilon_{\rm B}^{0.5}(E_{\rm iso}/4\times10^{54}\; {\rm erg})^{0.5}(\eta/0.1)^{-0.5}\\
&\;\;\;\;\;\;\;\;\;\;\;\;\;\;\;\;\;\;\;\;(t/200 \;{\rm s})^{-1.5}\;\; {\rm Hz} \\
 \nu_{\rm c} &\sim  2 \times 10^{12} \epsilon_{\rm B}^{-3/2}(E_{\rm iso}/4\times10^{54} \; {\rm erg})^{-0.5}(\eta/0.1)^{0.5}\\
 &\;\;\;\;\;\;\;\;\;\;\;\;\;\;\;\;\;\;\;\;(n_{\rm ISM}/ 0.3 \;{\rm \; {\rm cm^{-3}}})^{-1}(t/200\;{\rm s})^{-0.5}\;\; {\rm Hz} \\
t_{\rm c} &\sim  2 \times 10^{-8} \epsilon_{\rm B}^{-3}(E_{\rm iso}/4\times10^{54} \; {\rm erg})^{-1}(\eta/0.1)\\
&\;\;\;\;\;\;\;\;\;\;\;\;\;\;\;\;\;\;\;\;(n_{\rm ISM}/ 0.3 \; {\rm cm^{-3}})^{-2}(\nu_{\rm obs}/10^{15}\; {\rm Hz})^{-2} \;\;{\rm day} \nonumber
\end{split}
\label{eq:nu_m-nu_c-ISM}
\end{equation}
where $\epsilon_e$ and $\epsilon_B$ are the fractions of electron and magnetic energy transferred from the shock energy.

If the condition on the optical afterglow is (1) $\nu_{\rm m} < \nu_{\rm opt} < \nu_{\rm c}$, then the cooling frequency, $ \nu_{\rm c}$, of the afterglow should never cross the optical band because no significant temporal break was observed during the optical observation.
 Thus, it requires $t_{\rm c}$ $>$ 1 day ($\sim$10$^5$ s), leading to $\epsilon_{\rm B}$ $<$  3 $\times$ 10$^{-3}$. In this condition, the temporal index is expressed as 3(1-$p$)/4, and from the observed temporal index in the optical band  ($\alpha_{\rm AG}$ $\sim$ -1.6), the electron index can be estimated as $p$ $\sim$ 3.2. Furthermore, to satisfy the condition $\nu_{\rm m} < \nu_{\rm opt}$ at $T_0$ + 200 s, by using $\epsilon_{\rm B}$ $<$  3 $\times$ 10$^{-3}$ we obtained $\epsilon_{\rm e}$ $<$  2 $\times$ 10$^{-3}$.  Using the constraints, we can calculate the expected flux densities in the optical and GeV bands, where formulations by \cite{2002ApJ...568..820G} are used.
  If $\epsilon_{\rm B} \sim \epsilon_{\rm e} \sim$ 10$^{-3}$ is adopted, then the expected flux density at $T_{\rm 0}$ + 200 s is calculated as $F_{\rm \nu}$ $\sim$  10 mJy, which is consistent with the observed flux density ($\sim$10 mJy).
 However, using the same parameters, the flux density in the GeV band ($\sim$10$^{23}$ Hz) at $T_{\rm 0}$ + 200 s is calculated to be $F_{\rm \nu}$ $\sim$  10$^{-10}$ mJy, which is much smaller than the observed flux density ($\sim$10$^{-6}$ mJy).  This is because  $\nu_{\rm c}$ is expected to be low ($\nu_{\rm c}$ $\sim$ 10$^{16}$ Hz) with the adopted parameters and then the expected GeV flux density becomes very low, where the expected photon index above $\nu_{\rm c}$ is  $\Gamma_{\rm ph}$ = -2.6 from $p$ $\sim$ 3.2.
  Note that it is impossible to explain the observed GeV flux density when  $\epsilon_{\rm B}$ $<$ 3 $\times$ 10$^{-3}$ and $\epsilon_{\rm e}$ $<$ 2$\times$ 10$^{-3}$ by the synchrotron emission from the external shock propagating through the ISM.
 
 Next, the condition of (2) $\nu_{\rm m} <  \nu_{\rm c} < \nu_{\rm opt}$ (or $\nu_{\rm c} <  \nu_{\rm m} < \nu_{\rm opt}$) is considered. Here, the expected electron index is estimated to be $p \sim$ 2.8 from the temporal index observed in the optical band.   From the condition of $\nu_{\rm c}$ $<$  $\nu_{\rm opt}$, we obtain $\epsilon_{\rm B}$ $>$  0.02. 
 In addition, from the condition of $\nu_{\rm m} < \nu_{\rm opt}$, $\epsilon_{\rm e}$ $<$  2 $\times$ 10$^{-3}$ is obtained when adopting $\epsilon_{\rm B}$ =  0.02. If we appropriately choose  $\epsilon_{\rm e}$ = 3$\times$10$^{-4}$ and $\epsilon_{\rm B}$ = 0.02, the optical flux density should be $\sim$20 mJy, which is consistent with the observed flux density.
However, as is the case with (1) $\nu_{\rm m} < \nu_{\rm opt} < \nu_{\rm c}$, the flux density in the GeV band ($\sim$10$^{23}$ Hz) at $T_{\rm 0}$ + 200 s is calculated to be $F_{\rm \nu}$ $\sim$ 10$^{-10}$ mJy, which is much smaller than the observed flux density of $\sim$10$^{-6}$ mJy. 
 Here, even if we take a higher $\epsilon_{\rm B}$  the observed GeV  flux cannot be  explained because $\nu_{\rm c}$ becomes low. 
 
  In both cases, the optical emission can be accurately interpreted as having an external-shock origin, while the GeV emission cannot be consistently explained by the same. Thus, another origin for the GeV emission must be discussed.
  While we have assumed $n_{\rm ISM}$ = 0.3 cm$^{-3}$ implied from $\Gamma_{\rm bulk}$ = 420,
the above qualitative conclusion is almost the same
for a higher $n_{\rm ISM}$ $\sim$ 500 cm$^{-3}$ using $\Gamma_{\rm bulk}$ = 170.
 
 \subsubsection{Wind case}
For a wind profile of the ambient medium, with $A^*$ =  1.5 when we use $\Gamma_{\rm bulk, min}$ = 170,  the deceleration time of the GRB ejecta can explain the optical onset at $T_{\rm 0}$ + 200 s,  as mentioned in Sect. \ref{sec:optical_GeV_onset}.
 As is the case with the ISM, two cases of the optical emission, (1) $\nu_{\rm m} < \nu_{\rm opt} < \nu_{\rm c}$ and (2) $\nu_{\rm m} <  \nu_{\rm c} < \nu_{\rm opt}$ (or $\nu_{\rm c} <  \nu_{\rm m} < \nu_{\rm opt}$), are considered.
 Here, $\nu_{\rm m}$ and $\nu_{\rm c}$ and the crossing time $t_{\rm c}$   can be written as follows \citep{2000ApJ...536..195C}, 
\begin{equation}
\begin{split}
\nu_{\rm m} &\sim 1\times10^{21} \epsilon_{\rm e}^{2} \epsilon_{\rm B}^{1/2} (E_{\rm iso}/4\times10^{54}\; {\rm erg})^{1/2}(\eta/0.1)^{-1/2} \\
&\;\;\;\;\;\;\;\;\;\;\;\;\;\;\;\;\;\;\;\;(t/200\;{\rm s})^{-3/2}\;\; {\rm Hz} \\
\nu_{\rm c} &\sim  2\times10^{10} \epsilon_{\rm B}^{-3/2} (A^*/1.5)^{-2}(E_{\rm iso}/4\times10^{54}\;{\rm erg})^{1/2} \\
&\;\;\;\;\;\;\;\;\;\;\;\;\;\;\;\;\;\;\;\;(\eta/0.1)^{-1/2}(t/200\;{\rm s})^{1/2}\;\;{\rm Hz} \\
t_{\rm c} &\sim  7\times10^{6} \epsilon_{\rm B}^{3} (\nu_{\rm obs}/10^{15}\; {\rm Hz})^2 (A^*/ 1.5)^{4}\\
&\;\;\;\;\;\;\;\;\;\;\;\;\;\;\;\;\;\;\;\;(E_{\rm iso}/4\times10^{54}\;{\rm erg})^{-1}(\eta/0.1)\;\; {\rm day} \nonumber
\end{split}
\label{eq:nu_m-nu_c-wind}
\end{equation}

 For the case of (1) $\nu_{\rm m} < \nu_{\rm opt} < \nu_{\rm c}$,  $\epsilon_{\rm B} <$  7 $\times$ 10$^{-4}$ is obtained from the condition $\nu_{\rm opt} < \nu_{\rm c}$. In addition, from the condition $\nu_{\rm opt}> \nu_{\rm m}$ and $\epsilon_{\rm B} <$  7 $\times$ 10$^{-4}$, $\epsilon_{\rm e} <$  10$^{-2}$ is obtained. To explain the observed flux density of $\sim$10 mJy at $T_{\rm 0}$ + 200 s in the optical band, $\epsilon_{\rm B} \sim$  10$^{-4}$ and $\epsilon_{\rm e} \sim$  10$^{-3}$ is chosen. However, when these parameters are used, the expected GeV flux density  of $F_{\rm \nu}$ $\sim$ 3$\times$10$^{-8}$ mJy at $T_{\rm 0}$ + 200 s is much lower than the observed value ($\sim$10$^{-6}$ mJy). If a lower $\epsilon_{\rm B}$ is taken, then the GeV flux increases, but the optical flux decreases: in such a parameter space, it is hard for the external-shock scenario to explain both the optical and GeV fluxes.
 
  In the case of (2) $\nu_{\rm m} < \nu_{\rm c} < \nu_{\rm opt}$ (or $\nu_{\rm c} < \nu_{\rm m} < \nu_{\rm opt}$),  since this requires $t_{\rm c}$   $>$ 1 day ($\sim$10$^5$ s), $\epsilon_{\rm B} >$  5 $\times$ 10$^{-3}$ is obtained.
Furthermore, from the condition of $\nu_{\rm opt} > \nu_{\rm m}$, $\epsilon_{\rm e} <  $  2$\times$10$^{-3}$ ($\epsilon_{\rm B}$/10$^{-2}$)$^{-1/4}$ is obtained. If $\epsilon_{\rm B} \sim$    10$^{-2}$ and $\epsilon_{\rm e} \sim$  3$\times$10$^{-4}$, then the expected flux in the optical band is roughly consistent with the observed flux ($\sim$10 mJy). However, in any parameter space within this constraint, a GeV flux consistent with the observed flux can not be found. 
Thus, it can be concluded that the optical afterglow can be interpreted within the external-shock model, while the GeV extended emission cannot because it has a different origin.

On the other hand, the GeV onset at $T_{\rm 0}$ + 20 s could be explained by the deceleration onset of the GRB ejecta when adopting $A^*$ =  20, as mentioned in Sect. \ref{sec:optical_GeV_onset}. Thus, the plausibility of using the framework of the external-shock scenario to explain the optical and GeV emissions was tested. In this case, the most possible scenario could be that the optical emission until $T_{\rm 0}$ + 200 s comes from a condition of $\nu_{\rm opt} < \nu_{\rm m} < \nu_{\rm c}$, and the expected temporal index is $\alpha_{\rm AG}$ = +0 \citep{1998ApJ...497L..17S}, which is consistent with the observed temporal index $\alpha_{\rm AG}$ = +0.10 $\pm$ 0.13 from $T_{\rm 0}$ + 150 s to 300 s.
This condition requires  $\nu_{\rm opt}$  $\sim$ $\nu_{\rm m}$  at $T_{\rm 0}$ + 200 s, which leads to 
$\epsilon_{\rm e}^{2}\epsilon_{\rm B}^{1/2} \sim$ 10$^{-6}$. Furthermore,  from the condition of $\nu_{\rm opt} < \nu_{\rm c}$, $\epsilon_{\rm B} < $   2$\times$10$^{-5}$ should be satisfied. When $\epsilon_{\rm e} \sim$  0.18 and $\epsilon_{\rm B}$ $\sim$  10$^{-9}$ are chosen, the optical flux is obtained as $F_{\rm \nu}$ $\sim$ 10 mJy at $T_{\rm 0}$ + 200 s.
 In addition, the GeV flux using these obtained physical parameters, $\epsilon_{\rm e}$ and $\epsilon_{\rm B}$, was estimated to be   $\sim$5$\times$10$^{-4}$ mJy  at $T_{\rm 0}$ + 20 s, which is  inconsistent with the observed flux of (0.8$\pm$0.4)$\times$10$^{-5}$ mJy. 
  If a higher  $\epsilon_{\rm B}$ is taken, the observed GeV flux can be realized but the calculated flux in the optical band becomes much larger than the observed one.
Thus,  when considering the wind case with $\nu_{\rm m} < \nu_{\rm c} < \nu_{\rm opt}$ (or $\nu_{\rm c} < \nu_{\rm m} < \nu_{\rm opt}$), the GeV onset  cannot be interpreted as the deceleration of external shock. 
 
 Furthermore, there are several difficulties in that the external-shock scenario cannot fully explain other observational features: the electron index from the observed temporal index in the optical band is calculated as $p \sim$ 2.5. When the obtained electron index is used, the temporal index under $\nu_{\rm c} < \nu_{\rm GeV}$ should give $\alpha_{\rm AG}$ = -1.38, which is slightly inconsistent with the observed temporal index ($\alpha_{\rm AG}$ = -1.05 $\pm$ 0.20) of the temporally extended gamma-ray emission.  In addition, the extended emission in the GeV band shows signs of strong temporal and spectral variability, as shown in Fig. \ref{lcExtendedEmission}, which cannot be easily explained by the external forward shock model. Thus, the external-shock scenario faces several challenges when explaining the GeV emission. 

  Instead, even if we use $\Gamma_{\rm bulk, max}$ $\sim$  420 for the wind scenario, we obtain an almost similar conclusion as in the case of $\Gamma_{\rm bulk, min} \sim$   170, as shown briefly in the following: if $A^*$ $\sim$  0.04,  while the optical onset can be interpreted as the deceleration timescale, the expected flux of the GeV extended emission from the external-shock model cannot be reconciled with the observed one either in the case of $\nu_{\rm m} < \nu_{\rm opt} < \nu_{\rm c}$,  $\nu_{\rm m} <  \nu_{\rm c} < \nu_{\rm opt}$ or $\nu_{\rm c} <  \nu_{\rm m} < \nu_{\rm opt}$. 
In addition, if $A^*$ $\sim$  0.4, the GeV and optical onsets can be explained by the deceleration timescale and the crossing time of $\nu_{\rm m}$ in the case of  $\nu_{\rm opt} < \nu_{\rm m} < \nu_{\rm c}$, respectively, and the optical and GeV fluxes can be  barely explained by the external-shock model. However, there are some observed features such as the temporal index in the GeV band and the GeV extended emission in time interval ``I\hspace{-.1em}I" showing strong temporal and spectral variability, which are inconsistent with ones expected from the external-shock model. 

Thus, our interpretation does not largely depend on the bulk Lorentz factor of the GRB ejecta as long as the bulk Lorentz factor has a fiducial value of a few hundred.

\subsection{Origin of GeV emission}
 A few of the previous works have shown that the temporal and spectral behaviors in the GeV band for the observed LAT GRBs can be described by synchrotron emission from the external forward shock scenario \citep{2010MNRAS.409..226K, 2010MNRAS.403..926G, 2010A&A...510L...7G}.
However, as mentioned in Sect.  \ref{subSec:AG_scenario},  
the external-shock model cannot adequately explain either the temporal or the spectral behavior measured from the multi-wavelength observations of GRB 170405A.

 A possibility to explain the GeV emission behavior could be to consider inverse Compton scattering: SSC emission from the forward or reverse shock,  or inverse Compton emission from the $e^{+/-}$ pair loaded blast wave \citep{2014ApJ...788...36B}.
  For SSC emission in the forward shock, e.g., in the case of $\nu_{\rm m} < \nu_{\rm opt} < \nu_{\rm c}$ of the  wind case discussed in Sect. \ref{subSec:AG_scenario}, low magnetization (i.e., $\epsilon_{\rm B}$ $\sim$  10$^{-4}$) may enhance the SSC component in the gamma-ray band. However, the component is expected to be subdominant to the synchrotron emission from the external shock \citep{2015MNRAS.453.3772L,2017ApJ...844...92F}. 
  In addition, because the SSC emission cannot have a longer delay than the variability timescale \citep{2010arXiv1003.2452G}, the observed delayed onset cannot be realized; the timescale of the delayed onset is $\sim$20 s and the variability timescale is $\sim$0.5 s in this GRB.
For SSC emission in a reverse shock, although a bright spike with a long delay in the GeV band can be described \citep{2016ApJ...831...22F, 2017ApJ...848...94F}, it is difficult to explain through this model the observed temporal  and spectral behaviors in the GeV band, showing signs of strong temporal variability  and spectral hardening particularly in the time interval  ``I\hspace{-.1em}I\hspace{-.1em}I". 

 Here,  inverse Compton emission from the pair-loaded blast wave could be an attractive model that can realize the spectral hardening in the later phase \citep{2014ApJ...789L..37V,2015ApJ...813...63H}, because  the average energy of the pairs increases with time owing to the decrease of the pairs. However, this model expects the same temporal onset in the optical and GeV band \citep{2014ApJ...789L..37V,2015ApJ...813...63H}, and it might be difficult to fully explain the observed temporal behavior for this GRB:  $T_{\rm 0}$ + $\sim$200 s in the optical band and  $T_{\rm 0}$ + $\sim$20 s in the GeV band.
  
 Another plausible scenario could be synchrotron emission from an internal shock, which can naturally explain the strong temporal variability in the GeV band. In addition, the delayed onset can be interpreted as strong temporal dependence of the pair-production opacity, which can be longer than the variability timescale \citep{2012MNRAS.421..525H}. To explain the observed behaviors in the GeV band for other LAT GRBs, the preference of an internal-shock scenario over other models was also suggested by \cite{2011MNRAS.415...77M}.

Most of the temporally extended GeV emission is consistent with
a single power-law decay. The later part of the GeV afterglows would
be explained by the standard synchrotron or inverse Compton emission
from the external shock. If the initial part of the GeV emission has
another origin as we have discussed, the smooth behavior of the GeV lightcurves
could raise a question of the relation between the two components,
though the photon statistics is not significant in many cases.

\section{Conclusion}\label{sec:conclusion}
This study has presented the spectral and temporal analysis of GRB 170405A with optical, X-ray, and gamma-ray data.
 During the prompt emission phase, GRB 170405A shows a delayed onset of gamma-ray emission with respect to the X-ray emission by $\sim$20 s, after which temporally extended emission lasting $\sim$1,000 s in the GeV band was detected.
  In addition, $\sim$200 s after the prompt emission started, a clear optical onset was observed by {\it Swift}/UVOT.
  If the optical onset was caused by the external shock, the GeV emission cannot be explained by the external-shock scenario in the context of the standard synchrotron shock model.
  In particular, a significant spectral cutoff at $\sim$50 MeV in the prompt emission was detected, which enabled the estimation of the bulk Lorentz factor of GRB ejecta,  assuming that this spectral cutoff is caused by pair-production opacity.  The initial bulk Lorentz factor was derived to be $\Gamma_{\rm bulk}$ = 170--420 and  we test the plausibility of the GeV or optical onsets being the deceleration onset expected from the external-shock scenario.
  Investigating both ISM and wind cases of the afterglow, the most reasonable scenario was found to be that the optical delayed onset was caused by the deceleration of the external shock.
  In this case, the external-shock model cannot synthetically explain both the temporal and spectral properties of the temporally extended emission in the GeV band, and another emission component is needed.
  
   Possible scenarios are the synchrotron emission from the internal shocks
or the inverse Compton emission from the pair-loaded blast wave.  The model with the pair-loaded blast wave  \citep{2014ApJ...788...36B} could be suitable  to explain the delayed temporal onset and the spectral hardening at the later phase in the GeV band, although we need to adjust the parameters to suppress the optical synchrotron emission
from the pair-loaded blast wave \citep{2014ApJ...789L..37V}, as the observed optical afterglow has a different onset time and may have
another origin.  

   For the internal-shock scenario, the delayed GeV emission can be naturally interpreted as strong temporal dependence of the pair-production opacity, which is supported by the detection of a spectral-cutoff feature in the high-energy band.
  Furthermore, the temporal variability observed in the GeV band can be also easily described.
Although the external-shock scenario in many cases has been used to explain the GeV emission behavior,  GRB 170405A is a good counterexample to revise the emission model in the GeV band, and these results of GRB 170405A will shed new light on our understanding of the gamma-ray emission mechanism of GRBs.

\acknowledgments
 We appreciate the anonymous reviewer for his/her valuable comments on our paper.
 
The \textit{Fermi} LAT Collaboration acknowledges generous ongoing support from a number of agencies and institutes that have supported both the development and the operation of the LAT as well as scientific data analysis. These include the National Aeronautics and Space Administration and the Department of Energy in the United States, the Commissariat \`a l'Energie Atomique and the Centre National de la Recherche Scientifique / Institut National de Physique Nucl\'eaire et de Physique des Particules in France, the Agenzia Spaziale Italiana and the Istituto Nazionale di Fisica Nucleare in Italy, the Ministry of Education, Culture, Sports, Science and Technology (MEXT), High Energy Accelerator Research Organization (KEK) and Japan Aerospace Exploration Agency (JAXA) in Japan, and  the K.~A.~Wallenberg Foundation, the Swedish Research Council and the Swedish National Space Board in Sweden. Additional support for science analysis during the operations phase is gratefully acknowledged from the Istituto Nazionale di Astrofisica in Italy and the Centre National d'\'Etudes Spatiales in France.

This work was supported by JSPS KAKENHI Grant Numbers JP17H06362 (M.A.), JP16J05742 (Y.T.),  the JSPS Leading Initiative for Excellent Young Researchers program (M.A.), Sakigake 2018 Project of Kanazawa University (M.A.), and the joint research program of the Institute for Cosmic Ray Research (ICRR), The University of Tokyo (K.A.).







\appendix

\clearpage

\end{document}